# Establishing Role-based Access Control in Viewpoint-oriented Variability Management

Tobias Kaufmann, Thorsten Weyer

paluno – The Ruhr Institute for Software Technology
Gerlingstraße 16, 45127 Essen, Germany
{tobias.kaufmann, thorsten.weyer}@paluno.uni-due.de

**Abstract.** Process roles are used to structure complex engineering processes in single systems development for many years. Typically, each role has specific responsibilities from which certain information demands originate. In the engineering of variable software, role-specific information demands affect variability information. To control the access to the variability information, we suggest using the concepts of an explicit access control model. We integrate an access control model and a variability modeling language on a conceptual level. Additionally, we extend the variability modeling language by formally defined operations. Based on this extension, we propose a formal integration of an access control model and the variability modeling language. Our solution allows to explicitly define access control to variability information on a formal basis.

## 1 Introduction

Process-roles (or short: roles) are used to structure complex engineering processes of single systems (cf. [1, 2]). Thus, role-specific responsibilities of the engineering process are managed. Based on these responsibilities, role-specific information demands w.r.t. to software engineering artifacts exist. These information demands are represented in role-specific information sets, containing a subset of engineering artifacts. Information sets are subject to concurrency conflicts because they can overlap. Not every role has the same permissions on the elements of an information set, so explicitly controlling the access to information sets is an accepted best practice to avoid such conflicts. Hence, established access control models are applied to the engineering artifacts. In this setting two needs for structuring an engineering process can be formulated: *Information sets need to be defined to meet the information demands of roles* (*Need 1*); *Permissions on elements and operations of information sets need to be managed* (*Need 2*)

Defining and managing information sets as well as controlling the access to their elements is also relevant in the engineering of variable software that is used today (cf. [3, 4]). Variable software enables different software products with different features, sharing a common core. Such software is typically developed for multiple markets and a heterogenic stakeholder base. In this setting, potential mutually exclusive stakeholder demands need to be considered. Thus, the first class concept of variability (cf. [3]) is used to explicitly document this crosscutting property in separate models. Variability is understood as a separate layer of information further increasing the complexity of an engineering process. Hence, variability management is key for engineering variable software. Assuming that variability information can be managed the same way as engineering artifacts, we understand that role-specific information sets and permissions for operations on role-specific information sets need to be specified.

To define information sets on the variability information layer, variability models need to be taken into account. Ways to generally define information sets on variability models were studied in the literature. CZARNECKI et al. [5] use different stages to enable a staged configuration of variability. Each stage represents a subset of variability information. MANNION et al. [6] propose stakeholder-specific subsets of a master variability model to represent different information sets. ACHER et al. [7] and HUBAUX et al. [8] realize information sets on variability models as views by projections. Hence, these approaches address *Need 1* and can be used to structure the variability information in an engineering process. These works do not focus on controlling the access to the elements of variability information sets. Therefore, these approaches cannot be applied in an engineering process that requires controlling the access to elements of variability information sets without modification. Consequently, *Need 2* is not addressed. Therefore, we address the problem that the sole use of views (variability information sets) is not sufficient to realize access control on variability models.

Our contribution enables the application of access control concepts directly on Orthogonal Variability Model (OVM, [3]) instances. Therefore, we first integrate the Role-based Access Control Model (RBAC, [9]) and OVM on a conceptual level. Based on this conceptual integration, we extend the OVM by formally defined operations. This extension allows the assignment of permissions on the operations applicable to OVM elements. These permissions can be defined in an explicit access control model instance that is independent of the OVM. Because the proposed integration is based on OVM element and operation level, the results of our work can be used to seamlessly apply the concepts of existing approaches for defining information sets on variability information (e.g. [5, 6, 7, 8]) with our approach. The extension of OVM and the integration of RBAC and OVM are necessary steps to address *Need 2*.

The paper is structured as follows: In Section 2, we present the fundamentals of our approach. Section 3 describes the ontology and requirements of our approach. Section 4 integrates RBAC and OVM and the viewpoint concept. In Section 5, we present an application example. Related work is presented in Section 6. Conclusion and future work are presented in Section 7.

## 2 Fundamentals

**Modeling Variability with Orthogonal Variability Model (OVM).** The concept of variability defines a first class concept that needs to be documented explicitly in a separate variability model (cf. [3, 4]). Its fundamental ontological concepts are variability subject and variability object. The variability subject is defined as a variable item of the real world or a variable property of such an item (e.g., car paint). Accordingly, the *variability object* is defined as a particular instance of a variability subject (e.g., red paint, cf. [3]). These two concepts are supported by multiple relationships between variability subjects and variability objects, such as optional, mandatory, or alternative group dependencies. Constraints between variability subjects and variability objects are used to express *requires-* or *excludes-* relationships. These concepts can be instantiated by different variability modeling languages (VML). The OVM documents variability information of variable software. OVM instantiates the concept variability subject and variability object as *variation point* and *variant*. The OVM supports uni-directional *requires* and bi-directional *excludes* concerns. Dependencies that express an optional or mandatory relation between a variation point and a variant are defined, as well as alternative groups (concrete Syntax: Fig. 1). Variants are always part of a variability dependency and are related to a variation point.

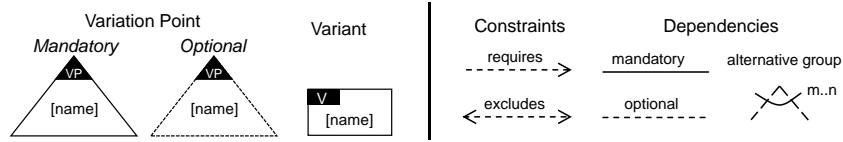

**Fig. 1.** OVM Concrete Syntax (cf. [3])

**Role-based Access Control (RBAC).** RBAC is an access control model standardized in ANSI/INCITS Std. 359-2004 [9]. This access control model consists of multiple sets. The core elements of RBAC are the sets *USERS*, *ROLES, OBJECTS, OPERATIONS*, and *PERMISSIONS*. In RBAC, a *user* is defined as a human being. A *role* is a job function within an organization and can hence be understood as a process role. Members of the set *OBJECTS* are defined as resources that are subject to access control. Elements of *OPERATIONS* are defined as executable images of a program that execute a function for the user who invoked the operation. A *permission* is defined as an approval to perform an operation on an object that is subject to access control. Roles are assigned to users (UA ⊆ U × R). Permissions are assigned to roles (PA ⊆ P × R). The RBAC reference model supports multiple extensions, such as role hierarchies and separation of duty. Fig. 2 shows the conceptual relations between the RBAC elements.

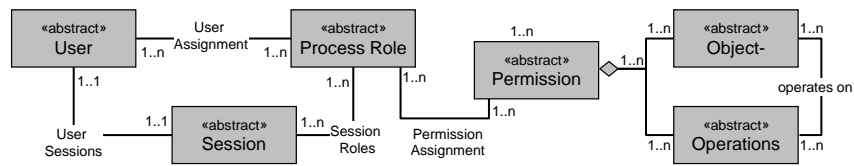

**Fig. 2.** Role-Based Access Control Conceptual Model (based on [9])

**IEEE Std. 42010-based Viewpoints.** The IEEE Std. 42010 [10] provides a conceptual framework for defining viewpoints. These viewpoints allow a separation of concerns in an architectural description. Concerns are the core concept of viewpoint definitions, describing the interest of stakeholders in the architecture of the software under development (SUD). A viewpoint is a specification supporting the structured derivation of one view on a SUD. Hence, multiple viewpoints are required to fully describe the architecture of such software.

## 3 Ontological Assumptions and Requirements of the Approach

Based on the concepts explained in Section 2, we describe the interplay of the fundamentals on an ontological level. We understand that in variability management, an access control model must integrate with existing concepts such as process roles (or short: roles), viewpoints and the concepts provided by a variability modeling language. Fig. 3 shows the ontology for integrating OVM, RBAC and Viewpoint-oriented concepts.

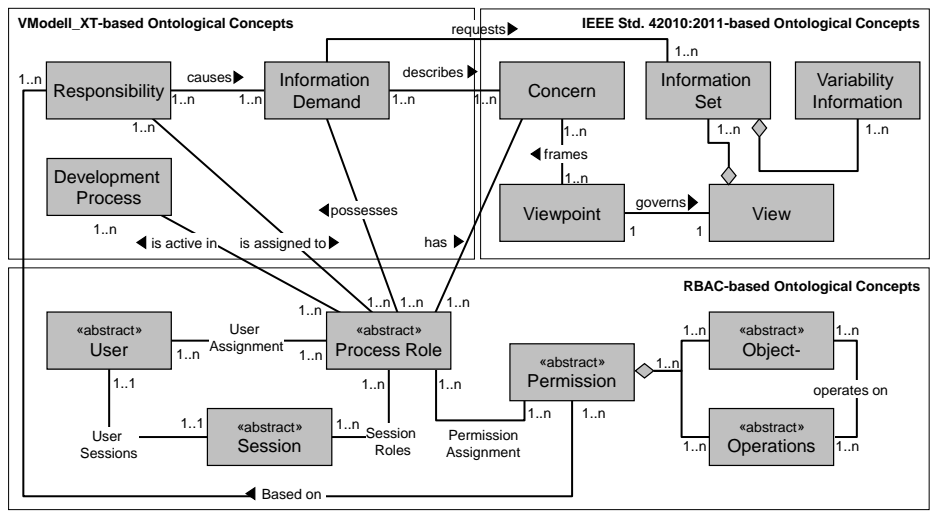

**Fig. 3.** Ontology for RBAC in Viewpoint-oriented Variability Management

Roles need to be recognized, because they are essential to established software process models (cf. [1]). Hence, roles are the connective element of our ontology. Roles have multiple diverging permissions, allowing them to execute certain operations on different objects. These role-specific permissions are based on the responsibilities that are assigned to a role, whereas the assignment of responsibilities is organized in the software process model. Roles have specific concerns that are documented in viewpoints. Based on the viewpoints, views can be derived that contain a role-specific information set (subset) of the complete set of variability information.

Roles are not disjunctive entities with regards to their information demands. Hence, information demands may be conflicting. This situation is known in the engineering of single systems but is commonly understood in the engineering of variable software.

Hence, we derive the following requirements for our approach:

- A variability model needs to be conceptually specified. **(R1)**
- The operations of a variability model need to be specified conceptually. **(R2)**
- Access to variability information needs to be explicitly controlled. **(R3)**
- Access control needs to fine granular. **(R4)**
- Permissions need to be assigned to process-roles. **(R5)**

The requirement R1 states that the elements of a variability model and the relations among them need to be made explicit in a conceptual model. Such a conceptual model can be understood as a meta-model (cf. [3]). Accordingly, requirement R2 states that the operations of a variability model need to be defined on the same level as the elements. The requirement R3 states that access to variability information needs to be explicitly controlled, for instance by an access control model. The requirement R4 states that access rights need to be specifiable on an artifact element level. To realize a fine granular access control, the operations manipulating elements of an artifact need to be recognized as well.

Requirement R4 implies that the operations on elements of a VML and the elements themselves need to be formally specified. Hence, the requirements R6 and R7 can be formulated.

- The elements of a VML need to be formally specified. **(R6)**
- The operations of a VML need to be formally specified. **(R7)**

Addressing requirement R6 and R7 is language-specific. For feature models, multiple works already addressed this topic (cf. [11, 12, 13, 14]). In contrast, solely the elements of the OVM are defined by the abstract syntax (R1, [3]). The semantics of the elements of the OVM are defined in METZGER et al. [15]. Thus, to the best of our knowledge, operations that manipulate an OVM are not yet formally specified.

The aforementioned requirements R3 – R7 aim at the integration of a variability model and an explicit access control model. In this setting, the need for a concept that is capable of defining role-specific information sets is also present, which can be formulated as requirement R8:

- An access controlled VML needs to be integrated with a viewpoint concept. **(R8)**

## 4   Access Control for Variability Information

In our experience, a holistic approach for variability management needs to regard three distinct layers. The first layer contains the variability information that is explicitly documented by use of a variability model. In the second layer, the relationships between the variability model and the base-artifacts of the engineering process are managed. The third layer contains all base-artifacts (e.g. requirements, components, classes, test cases) that can be related to variability information. In this paper, we regard the first layer in such a holistic approach towards variability management in order to integrate access control and variability modeling on a formally sound basis. Because RBAC is specified in Z, we also chose Z [16] in order to gain a sound integration of the concepts. Fig. 4 shows the hierarchical structure and the corresponding parts of our solution.

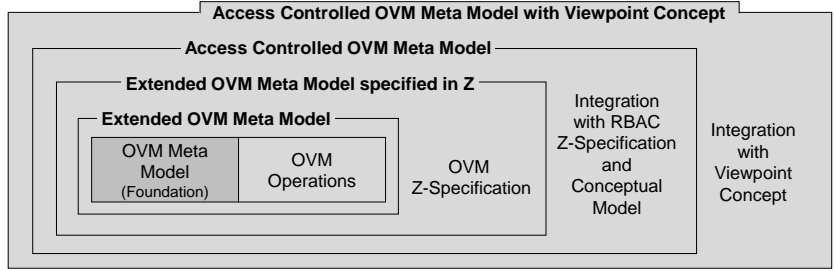

**Fig. 4.** Overview of the Structure of the Solution Approach

Each solution part addresses certain requirements stated in Section 3. Our solution consists of five parts. The first part "OVM Meta Model" has already been published and provides the basis for the solution presented in this paper. We describe the remaining solution parts in detail in the following sections.

- *Already existing OVM meta model* ($\rightarrow$ **R1**)  (Section 2)
- Extended OVM meta model ($\rightarrow$ **R2**)  (Section 4.1)
- Z-specification of the extended OVM ($\rightarrow$ **R6**, **R7**)  (Section 4.2)
- Integration of OVM and RBAC ($\rightarrow$ **R3**, **R4**, **R5**)  (Section 4.3)
- Integration with viewpoint concept ($\rightarrow$ **R8**)  (Section 4.4)

### 4.1  Extended OVM Meta Model

The OVM meta-model is presented in (cf. [3]). This conceptual model describes the elements of an OVM model instance as the abstract meta-classes Variation Point, Variant, Variability Dependency, and multiple constraint dependencies. Thereby, the abstract syntax of the OVM is described. In order to represent operations that can be applied to OVM elements, the original OVM meta-model was extended by abstract meta-class OVM Operations (cf. Fig. 8, Extended OVM Meta Model). An OVM operation can operate on the meta-classes Variation Point, Variant, Variability Dependency, and the constraint dependencies but is not limited to one element kind. By introducing OVM operations on the conceptual level, we address requirement R2.

### 4.2  OVM Z-Specification

OVM cannot be integrated with RBAC without formally defining operations for this variability modeling language. As RBAC is specified in *Z*, the OVM and the operations on concrete OVMs need to be defined in Z as well. The formal semantics of the OVM are presented in [15]. To specify operations that can be executed on a concrete OVM, we describe 'VP' and 'V' as the set of all possible variation points and variants. We further assume that the total amount of elements of V is equal to amount of integer numbers ($|V| \leq |\mathbb{N}|$). Furthermore, we assume the total amount of variation points to be less or equal to the total amount of variants. Hence, the $|VP| \leq |V| \leq |\mathbb{N}|$ is considered as always fulfilled. Each element of 'V' and 'VP' has a unique name. In a Z-based specification the sets 'VP' and 'V' can be represented as global variables. Because of the conceptual differences between variation points and variants, constraints need to be defined on a fine granular level.

```
1    ___OVM_Elements___                          ___OVM_Dependencies___
2    MAN_VP, OPT_VP: ℙVP                         MAN, OPT: V → VP
3    VARIANT: ℙV                                 ALTGROUP: (ℙ V×(N×N)) → VP
4    ∀x, y: VP • x ∈ MAN_VP ⇒ x ∉ OPT_VP ∧       ∀x, y: (V × VP) • x ∈ MAN ⇒ x ∉ OPT ∧
5      y ∈ OPT_VP ⇒ y ∉ MAN_VP                     y ∈ OPT ⇒ y ∉ MAN
6                                                ∃ x:VP; v: ℙ V; w: V; m,n:N | m ≤ n ≤ #v ∧
7                                                  w ∈ v ⇒ (w ↦ x) ∉ MAN ∧ (w ↦ x) ∉ OPT
8                                                  • (v ↦ (m↦n)↦ x) ∈ ALTGROUP
9
```

**Fig. 5.** Z-schemata OVM_Elements and OVM Dependencies

The Z-schema "OVM_Elements" declares three power sets (denoted by ℙ, cf. Fig. 5, lines 2, and 3) representing concrete mandatory and optional variation points, as well as concrete variants of an OVM model. The properties of these sets are defined in the lines 5 - 10. In Fig. 5, the sets share the property that their elements are unique. Three sets are declared in the first 2 lines of the schema "OVM_Dependencies". These sets represent the different variability dependencies (cf. OVM meta model, [3]) as surjective functions. The set 'ALTGROUP' describes concrete alternative groups consisting of two or more variants, as well as a minimal and a maximal cardinality that are related to a variation point (cf. Fig. 1). Each variant of an alternative group must not be a member of any other variability dependency (lines 7 - 9). The Z-schema "OVM_Constraints" represents the possible combinations of constraints between variation points and variants as separate sets (Fig. 6). Each set has a specified type in order to express the mapping of the elements of the set. Furthermore, there are no subset relations defined for type compatible constraint sets. Hence, we understand that elements of one constraint set cannot be member of another constraint set, even if these sets are type compatible. For instance, elements of the set 'EXCLUDES_V_V' can be described as follows: Let $v_1$, $v_2$, $v_3$, and $v_4$ be elements of type V. Furthermore, $v_1$, $v_2$, $v_3$, and $v_4$ are elements of 'VARIANT' (cf. Z-schema "OVM_Elements"). If there exists one excludes constraint between each variant pair ($v_1$, $v_2$) and ($v_3$, $v_4$), one has 'EXCLUDES_V_V' = {$v_1 \mapsto v_2$, $v_2 \mapsto v_1$, $v_3 \mapsto v_4$, $v_4 \mapsto v_3$}.

```
1    ___OVM_Constraints___                       ___add_Variation_Point___
2    EXCLUDES_V_V: V ↔ V                         ΔOVM_Elements
3    EXCLUDES_V_VP: V ↔ VP                       addManVP!: VP → ℙ VP
4    EXCLUDES_VP_V: VP ↔ V                       x?:VP
5    EXCLUDES_VP_VP: VP ↔ VP                     x? ∉ MAN_VP ∧ x? ∉ OPT_VP ∧ MAN_VP ⊂
6    REQUIRES_V_V: V ↔ V                          (MAN_VP ∪ {x?}) ∧ addManVP!(x?) =
7    REQUIRES_V_VP: V ↔ VP                        MAN_VP ∪ {x?}
8    REQUIRES_VP_V: VP ↔ V
9    REQUIRES_VP_VP: VP ↔ VP                     ___OVM_Model___
10                                               ΞOVM_Elements
11                                               ΞOVM_Dependencies
12                                               ΞOVM_Constraints
```

**Fig. 6.** Z-schemata OVM_Constraints, OVM_Model, add_Variation_Point

It is important to note that the elements of 'EXCLUDES_V_V', ($v_1$, $v_2$) and ($v_2$, $v_1$) are considered different. Furthermore, both elements are part of this excludes set, because excludes is a bi-directional constraint. For the sake of brevity, we omitted the explicit specification of uniqueness of the elements of each constraint set. The Z-schema "add_Variation_Point" specifies the operations *addManVP* (cf. Fig. 6). This operation takes a mandatory variation point as input and adds this variation point to an OVM model by uniting the existing set 'MAN_VP' with the set representation of the input parameter vp. The lines 5 - 7 of Fig. 6 state the properties that need to be fulfilled for these operations. Because of the uniqueness of elements in an OVM model, we explicitly state that the variation point to be added must not already be contained in the set 'MAN_VP'. Furthermore, we explicitly state that the unmodified set 'MAN_VP' is a proper subset of its modified version. The same operation can be defined to add optional variation points but was omitted for the sake of brevity in this paper. The Z-schema "OVM_Model" (Fig. 6) represents an OVM model by combining the aforementioned schemata. Hence, there are no further declarations or properties specified as they are specified in the corresponding schemata. No modifications to the sets of the referenced schemata are specified in "OVM_Model". To explicitly denote that the states of the referenced schemata are not changed, they are prefixed with the symbol Ξ. The Z-schema "remove_Variation_Point" specifies an operation that can be used to remove a variation point from the set 'MAN_VP'. Thus, the properties that must be satisfied are made explicit in the lines 7 - 13 in Fig. 7 for the operation *removeManVP*. This operation requires a variation point as input and removes it from the set 'MAN_VP', if and only if the variation point to be removed is an element of 'MAN_VP' and is not part of a constraint relation. Furthermore, the variation point to be removed must not be part of a variability dependency. The same operation can be defined to remove elements of the set 'OPT_VP'. This operation shares the same properties as the operation *removeManVP*. Due to space limitations, we present an excerpt of the operations. Nevertheless, this excerpt addresses the requirements R6 and R7.

```
1     remove_Variation_Point
2     ΔOVM_Elements
3     ΞOVM_Dependencies
4     ΞOVM_Constraints
5     removeManVP!: VP → ℙVP
6     x?:VP
7     ∀y:VP; v: V; m,n : ℕ; w:ℙV | x? ≠ y ⋀ m ≤ n ≤ #w ⋀ x? ∈ MAN_VP ⋀
8        ((w ↦ (m↦n)) ↦ x?) ∉ ALTGROUP ⋀ (v ↦ x?) ∉ MAN ⋀ (v ↦ x?) ∉ OPT ⋀
9        (x? ↦ y) ∉ EXCLUDES_VP_VP ⋀ (y ↦ x?) ∉ EXCLUDES_VP_VP ⋀
10       (x? ↦ y) ∉ REQUIRES_VP_VP ⋀ (y ↦ x?) ∉ REQUIRES_VP_VP ⋀
11       (v ↦ x?) ∉ EXCLUDES_V_VP ⋀ (x? ↦ v) ∉ EXCLUDES_VP_V ⋀
12       (v ↦ x?) ∉ REQUIRES_V_VP ⋀ (x? ↦ v) ∉ REQUIRES_VP_V •
13       removeManVP!(x?)= MAN_VP \ {x?}
```

**Fig. 7.** Z-schema remove_Variation_Point

### 4.3 Access Controlled OVM Meta Model

OVM and RBAC can be integrated on a conceptual level. This integration is based on the ontology (Section 3) and the extended OVM meta-model (cf. Fig. 8). Key to the proposed integration of RBAC and OVM is the understanding of Information Set and

Variability Information as specializations of Object (cf. Fig. 8). Moreover, Variability Information is in a specialization relation to Operations. Because of these specialization relations, Information Set and Variability Information are manageable by the concepts of RBAC. OVM Operations are in a specialization relation to Variability Information Operations (cf. Fig. 8). Because of the inheritance hierarchy, OVM Operations are RBAC Operations. The meta-classes Variation Point, Variant, Variability Dependency, and the different constraint dependencies are also understood as specializations of Variability Information. Because of the inheritance hierarchy these meta-classes are specializations of Object (cf. Fig. 8). This implies that the OVM meta-classes are either specializations of the Object or Operations. Consequently, OVM operations and the other OVM meta-classes can be managed using RBAC. Because of these relationships, OVM and RBAC are understood as conceptually integrated.

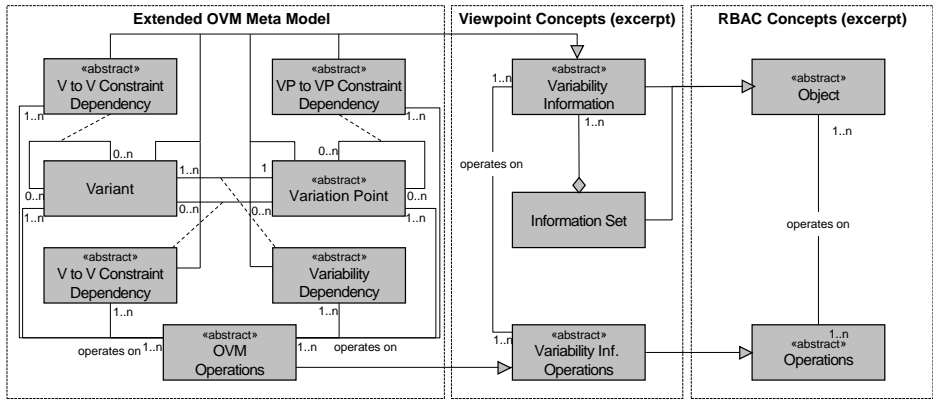

**Fig. 8.** Extended OVM Meta Model and integrated with RBAC

Integrating OVM and RBAC on the level of their Z-specifications, the elements and operations of the OVM Z-specification need to be related to their corresponding elements of the RBAC Z-specification. From this, it follows that we consider the sets of mandatory and optional variation points (cf. Fig. 5, Z-schema "OVM_Element") as disjoint proper subsets of the RBAC set OBJECTS. The same holds true for the sets defined in the Z-schema "OVM_Dependencies" (cf. Fig. 5), and "OVM_Constraints" (cf. Fig. 6). We integrate the previously defined operations in the same way by defining a proper subset relation between the operations *addManVP* (cf. Fig. 6), *removeManVP* (cf. Fig. 7) and the RBAC set 'OPERATIONS'. Hence, we consider the requirements R3, R4, and R5 addressed.

### 4.4 Access Controlled OVM Meta Model with Viewpoint-Concept

The conceptual integration of viewpoints/views and role-based access control is based on our ontology (Section 3). The ontology specifies the conceptual relation of a role-specific view and information set as an aggregation (cf. Section 3). Because of the specialization, the relation between information set and Object (RBAC) information set is considered integrated with RBAC. The same holds true for pieces of variability information that are the elements of a concrete information set. For this reason, we consider OVM and RBAC conceptually integrated with the viewpoint/view concept.

Defining role-specific views on the level of the Z-specifications is inherently realized by the assignment of permissions to roles. Because each permission consists of an object and an operation (cf. Fig. 2), views can be realised by making use of the RBAC operation *RolePermissions*. This operation is part of the RBAC [9] and requires a role as input parameter. If and only if the role is known, the objects and operations that are assigned to a role are returned. From this, it follows that the operation *RolePermissions* returns a subset of information, which in our case is an information set containing a role-specific subset of variability information. Based on the above explanations we consider requirement R8 addressed.

## 5 Application of our Approach

To illustrate our approach, we use the example of a product line from the medical domain published by Acher et al. [7].

### 5.1 OVM Example – Variability of Imaging Registration Service

Fig. 9 depicts an excerpt of the variability of the Medical Imaging Registration Service (or short: Imaging Registration Service) proposed in [7]. In [7] the variability of the Imaging Registration Service is documented in a feature model. Hence, we followed the approach described in [15] to extract the variability information from the feature model. We chose the variability model proposed in [7] because it explicitly refers to roles and uses views as role-specific information sets.

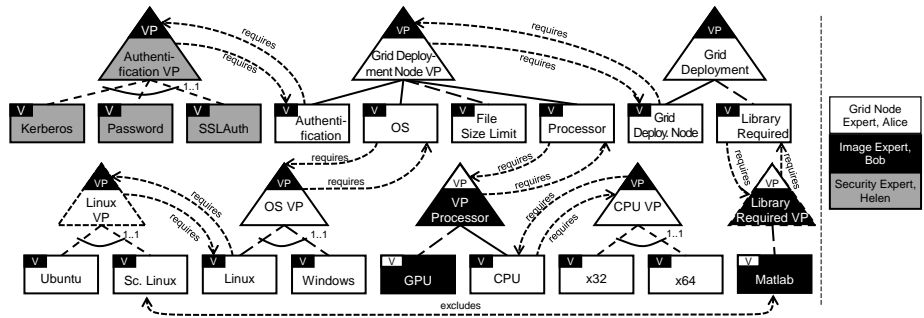

**Fig. 9.** OVM Example with role-specific views based on ACHER et al. (cf. [7])

### 5.2 Z-Specification – Variability of Imaging Registration Service

The example in Fig. 9 can be specified using the proposed Z-specification. Based on Fig. 7, we assign the elements of Fig. 9 to the sets of the Z-specification in Fig. 10. For the sake of brevity, we omitted the full specification of the example in this paper. A more detailed specification that contains the complete assignment of OVM elements from Fig. 9 to the elements of our Z-specification is available at http://goo.gl/FmwgdQ.

```
1   MAN_VP = {"Authentification VP", "Grid Deployment Node VP", "Grid Deployment", "OS VP",
2       "Processor VP", "CPU VP"}
3   OPT_VP = {"Linux VP", "Library Required VP"}
4   VARIANT = {"Kerberos", "Password", "SSLAuth", "Authentification", "OS", "File Size Limit", "Proces-
5       sor", …, "Sc. Linux", "Linux", "Windows", "GPU", "CPU", "x32", "x64", "Matlab"}
6   MAN = {"Authentification" ↦ "Grid Deployment Node VP", "OS" ↦ "Grid Deployment Node VP", …,
7       "Grid Deployment Node" ↦ "Grid Deployment", "CPU" ↦ "VP Processor"}
8   OPT = {"File Size Limit" ↦ "Grid Deployment Node VP", … "GPU" ↦ "VP Processor"}
9   ALTGROUP = {("Kerberos", "Password", "SSLAuth") ↦ (1,1) ↦ "Authentification VP"), ..,
10      ("x32", "x64") ↦ (1,1) ↦ ("CPU VP")}
11  REQUIRES_V_VP = {"Authentification" ↦ "Authentification VP", …, "Library Required" ↦ "Library Re-
12      quired VP", "Linux" ↦ "Linux VP", "OS" ↦ "OS VP", "CPU" ↦ "CPU VP"}
13  REQUIRES_VP_V = {"Authentification VP" ↦ "Authentification", " …, "Linux VP" ↦ "Linux",}
14  EXCLUDES_V_V= {"Matlab" ↦ "Sc.Linux", "Sc.Linux" ↦ "Matlab"}
```

**Fig. 10.** Z-Specification of OVM Example (excerpt)

### 5.3 Access Controlled Variability of Imaging Registration Service

To specify access to different objects using RBAC, we introduce NAME as an abstract data type. The elements of this data type are considered unique. Hence, we use a set of users specified in line 1 of Fig. 11. The set of roles is based on [7]. According to RBAC, the operation *AssignUser(user, role: NAME)* links a user to a specified role. In our example, the application of this operation can result in the user to role mapping specified in line 3 of Fig. 11. Due to space limitations, the members of the set 'OPERATIONS' contain the operations specified in Section 4.2. Further operations need to be defined to make full use of the specification.

In RBAC, a permission is a three-tuple of an object, an operation, and a role. Defining a permission in our setting requires to call the *GrantPermission2(objects: ℙNAME; operation, role: NAME)* operation. This is a slight modification to the original *GrantPermission* [9] operation, because it takes a set of objects as input instead of only one object per operation call. Hence, *GrantPermission2* creates an (object, operation) permission pair for each member of objects and assigns each pair to role, if and only if role is known to the RBAC system. In Fig. 12 the call to *GrantPermission2* in line 1 enables the role -Grid Node Expert- to read all elements of the set 'OBJECTS'. Hence, all elements of the variability model (Fig. 9) can be read. Furthermore, the role -Grid Node Expert- is allowed to add and remove mandatory variation points from the model. In our example, the user Alice has this role and can therefore execute these operations. In contrast, adding and removing optional variation points is not permitted for this role.

```
1   U = {"Alice", "Helen", "Bob"}
2   R = {"Grid Node Expert", "Image Expert", "Security Expert"}
3   UA = {("Grid Node Expert", "Alice"), ("Image Expert", "Helen"), ("Security Expert", "Bob")}
4   OPERATIONS = {addManVP,…, removeManVP }
5   OBJECTS = {MAN_VP, OPT_VP, VARIANT, MAN OPT, AlTGROUP, EXCLUDES_V_V,
6       EXCLUDES_V_VP, EXCLUDES_VP_V, EXCLUDES_VP_VP, REQUIRES_V_V,
7       REQUIRES_V_VP, REQUIRES_VP_V, REQUIRES_VP_VP }
```

**Fig. 11.** Specification of users, roles, operations applicable to OVM objects

### 5.4 Access Controlled Variability of Imaging Registration Service with Views

A view contains one or more role-specific information sets. In order to comply with this understanding of role-specific views, it is sufficient to retrieve a subset of the complete set of variability information. In our setting, executing *RolePermissions(role: NAME; result:2 $^{PERMS}$)* (cf. [9]) would result in a role-specific information set, if and only if the role is known and has permissions assigned. Because a permission aggregates an object and an operation that a role is allowed to execute on that object.

```
1   GrantPermission2(OBJECTS, read, "Grid Node Expert")
2   GrantPermission2(MAN_VP, add_Variation_Point, "Grid Node Expert")
3   GrantPermission2(MAN_VP, remove_Variation_Point, "Grid Node Expert")
4   GrantPermission2((("Kerberos","Password","SSLAuth")↦(1,1)↦("Authentification VP")), readAltGroup,
5       "Security Expert")
6   GrantPermission2((("Kerberos","Password","SSLAuth") ↦(1,1)↦("Authentification VP")),
7       writeAltGroup, "Security Expert")
8   GrantPermission2(("Matlab" ↦ "Library Required VP"), readOptDep, "Image Expert")
9   GrantPermission2(("Matlab" ↦ "Library Required VP"), writeOptDep, "Image Expert")
10  GrantPermission2(("GPU" ↦ "Processor VP"), read, "Image Expert")
```

**Fig. 12.** Specification of permissions based on objects, operations, and roles

The execution of the operation *RolePermissions* for the role -Grid Node Expert-, returns all permissions for that role, which are denoted on column two of Table 1. Apparently, the results denoted in column 2 (Result) would contain the proper (object, operation) pairs. We use ('OBJECTS', *read*) as an abbreviation for $(obj_1, read) \ldots (obj_n, read)$. The same applies to 'MAN_VP', *add/remove_Variation_Point*.

**Table 1.** Retrieval of role-specific information subset

|   | Call to Operation | Result |
|---|---|---|
| 1 | *RolePermissions("Grid Node Expert")* | (OBJECTS, read) |
| 2 | | (MAN_VP, add_Variation_Point) |
| 3 | | (MAN_VP, remove_Variation_Point) |

### 5.5 Example Discussion and Summary

Executing certain operations (e.g. *removeManVP*) on a concrete variability model (cf. Fig. 10) can have multiple side effects w.r.t. its well-formedness, as well as to its semantic correctness. To preserve the well-formedness of a concrete variability model, we specified certain preconditions for *removeManVP*. On the semantic level removing a variation point has wide-ranging consequences because removing a variation point is understood as narrowing the amount of configuration choices. Thus, a decision about the variants and constraints that are potentially related to his variation point needs to be made prior to its removal. Specific roles in an engineering process have the necessary responsibilities and the skill set to execute this removal operation. In our example a variation point can only be removed if it is not part of a variability dependency and a constraint. Consequently, a specific removal strategy is implied that may not be suitable for all roles. From this it follows that implementations of certain operations that manipulate a concrete variability model can also be role-specific. For instance if a product manager removes a variation point and the related variants, it is expected that this role is aware of the consequences. In contrast, if a technical engineer executes the same operation it is not guaranteed that he has assessed the consequences of this operation. This issue is not specific to *removeManVP*. Furthermore, this issue also applies to other operations and is variability modeling language independent.

We demonstrated that a concrete variability model can be represented by our Z-specification (Section 5.2). We illustrated how access control can be realized by applying the concepts of RBAC directly on the elements of the example variability model (Section 5.3). In addition, we pointed out how access controlled role-specific information sets (views) can be derived by applying RBAC operations (Section 5.4). The application of our approach showed that the requirements (Section 3) for establishing role-based access control in viewpoint-oriented variability management have been met.

## 6 Related Work

Multiple areas of research are relevant to our proposed approach. BATORY [17] uses grammars and propositional formulas to formally specify feature models. MANNION [18] uses first order logic to formally represent a product line model and to validate the product line. SCHOBBENS et al. [19] present an overview on different formalizations of feature models. SUN et al. [11] provide a Z-based specification for feature models. This specification documents the semantic aspects of the tree structure of feature models by recognizing different kind of parent and child relations between features. In contrast to our work, operations on features are not considered. A Z-specification for non-attributed feature models is also provided in YE and LIN [12]. YE and ZHANG [13] propose a Z-specification for software product line evolution. BENAVIDES uses a Z-based specification for attributed feature models in the layered FAMA-Framework [14] in order to reason about attributed feature models. These works do not make the connection between access control and the provided specification. Because of the conceptual difference between the used feature modeling languages and the OVM, the proposed Z-specifications cannot be reused in our setting. MUTHIG and SCHROETER [20] propose a controlling of access to feature information during engineering in software product line organizations. The work in [20] focusses on access control for feature information and not on variability information represented as features. Furthermore, this work demonstrates a conceptual integration of RBAC and feature information.

Multiple different access control models have been discussed and studied in the literature. LAMPSON [21] introduces the concept of an access control matrix that can be implemented by access control lists [22]. The access control matrix proposed in [21] does not explicitly recognize roles and is hence not suitable for our approach. Task-based authorization by THOMAS and SANDHU [23] focusses on data integrity from an organizational and sociological perspective. This access control model is too high level and thus not suitable for controlling access to variability information. A spatial approach to access control was proposed by BULLOCK et al [24]. Because of the coarse level of access control, this approach is not suitable without modification. RBAC is based on the work of FERRAIOLO et al. [25] and lays the foundation for other access control models. Team-based Access Control (TMAC) by THOMAS [26] is an application of RBAC in a collaborative environment. GEORGIADIS et al. [27] propose an extension of TMAC that uses context information to activate permissions.

## 7   Summary and Outlook

In this paper, we proposed a solution for the general problem of access control to variability information in a specific setting. Part of this setting is the Orthogonal Variability Model (OVM) and the Role-based Access Control (RBAC) model. Therefore, the OVM was integrated with RBAC on a conceptual and a formal level. Furthermore, it was demonstrated how views can be applied in a RBAC setting. The applicability of our approach to the existing body of knowledge was demonstrated by an example taken from the literature. Furthermore, we briefly discussed the potentially wide-ranging effects of operations on a variability model. Thereby, we demonstrated why access control of variability information is important for holistic variability management and hence for preventing unwanted change of variability information. Our approach can be generalized to support other variability modeling languages, such as the different feature model dialects discussed in [19]. In order to do so, a Z-specification for the selected feature modeling dialect is required that specifies the elements and the operations that are allowed on the corresponding feature models. Moreover, the relation between feature model elements and RBAC elements need to be defined, as well as the relation between RBAC operations and feature model operations.

This work is part of a broader research agenda. In the next step, we strive to extend our solution to the base-artefact to variability relation layer, as well as to the base-artifact layer. Thereby, we aim at a complete integration of access control and views across all three variability management layers. We also plan to investigate the challenges ahead that result from scenarios in which inconsistent access control information on the different layers are present. To further evaluate our proposed approach, we plan to implement our solution in our variability modeling tool proposed in [28]. The extended version of this tool is planned to be applied in an industrial setting with our partners from the German automotive industry.